\documentclass{ws-ijmpd}

\begin{document}
\def\be{\begin{equation}}
\def\ee{\end{equation}}

\markboth{J.E. Horvath and P.H.R.S. Moraes} {Modeling a $2.5 \, M_{\odot}$ compact star with quark matter}

%
\catchline{}{}{}{}{}
%

\title{MODELLING A $2.5 \, M_{\odot}$ COMPACT STAR WITH QUARK MATTER}

\author{J. E. HORVATH$\dag$}
\author{P. H. R. S. MORAES$\dag$}

\address{$\dag$Instituto de Astronomia, Geof\'\i sica e Ci\^encias Amosf\'ericas, Universidade de S\~ao Paulo,\\
S\~ao Paulo 05508-090/SP,
Brazil}

\maketitle

\begin{history}
\received{Day Month Year}
\revised{Day Month Year}
\end{history}

\begin{abstract}
The detection of an unexpected $\sim 2.5 M_{\odot}$ component in the gravitational wave event
GW190814 has puzzled the community of High-Energy astrophysicists, since in the absence of further information it is not clear whether this is the heaviest ``neutron star'' ever detected or either the lightest black hole known, of a kind absent in the local neighbourhood. We show in this work a few possibilities for a model of the former, in the framework of three different
quark matter models with and without anisotropy in the interior pressure. As representatives of classes of ``exotic'' solutions, we show that even though the stellar sequences may reach
this ballpark, it is difficult to fulfill simultaneously the constraint of the radius as
measured by the NICER team for the pulsar PSR J0030+0451. Thus, and assuming both measurements stand, compact neutron stars can not be all made of self-bound quark matter, even within anisotropic solutions which boost the maximum mass well above the $\sim 2.5 M_{\odot}$ figure. We also point out that a very massive compact star will limit the absolute maximum matter density in the present Universe to be less than 6 times the nuclear saturation value.
\end{abstract}


\section{Introduction}

Even though the idea of a ``canonical'' $1.4 M_{\odot}$ mass for compact stars has been around for years, it became clear in the last decade or so that this single-scale is not tenable \cite{HV}. At least one additional peak, and most likely two \cite{nos1} are present in the mass distribution, irrespective of the large uncertainties in many objects, particularly in those at the highest masses \cite{nos2}. The issue of a maximum compact star mass is also under discussion, since the latest accurate measurement of a compact object using Shapiro delay \cite{Cromartie} yielded $2.14{{+0.1}\atop{-0.09}} M_{\odot}$ for the millisecond pulsar MSP J0740+6620. A few higher measured values exist, although they have been extracted using methods not generally as reliable. Anyhow it has been argued that a handful of compact stars in the group of interacting ``redback/black widow'' systems may achieve masses greater than the MSP J0740+6620 \cite{Cromartie}, and possibly all the way up to the maximum TOV value, due to their evolutionary histories \cite{SciChi}. It is even possible to produce some ``low-mass'' black holes if some of these systems push the mass of the accreting pulsar over the TOV limit.

On the other hand, and as a valuable tool to complement the knowledge of compact stars in binaries, the availability of gravitational wave data has provided exciting news about the systems merging presently. While the celebrated GW170817 event \cite{abbott/2017} inaugurated the Multimessenger Era, being detected by more than 60 instruments around the world, a few other intriguing detections and candidates are available, although the merging rate does not appear to be as high as originally expected \cite{Luciano}.

One particular event denoted as GW190814 has shown a very intriguing component: an individual mass in the range $2.5-2.67 M_{\odot}$ ($90\%$ confidence) \cite{Abbott2020}. In the absence of additional information, it is not clear whether this is a light black hole or rather a very massive compact star (a cumbersome nickname of ``black neutron star'' \cite{BNS} was used in the media to reflect this ambiguity, although its precise meaning is not related to a ``hybrid'' character and may be misleading). Then, if the second possibility holds, it may be necessary to enlarge the maximum mass value to accommodate it. This in turn would be very important for the microphysics of the dense matter above the saturation density. Some works have already appeared discussing this possibility \cite{Ignazio,Xu}.

Motivated by this evidence, we revisit in this article a class of self-bound stellar models entertained in the last decade to provide viable stellar sequences complying with a high maximum mass. We shall  present these solutions below, together with a comparison with the recently published data about the radius of a ``lighter'' compact star, namely the pulsar PSR J0030+0451 \cite{Col}, the only case with a reliable determination of the radius at present. It will be shown that there is a quandary with an explanation of both data simultaneously. Finally, we point related additional information related to the maximum mass problem.

In our calculations we shall consider three different
quark matter models with and without anisotropy in the interior pressure. The quark matter models will be described in the next section. For now it is worth mentioning that anisotropy may arise naturally in matter fields at high densities ($\rho>10^{15}$g/cm$^3$) \cite{ruderman/1972,canuto/1974} and play a fundamental role in the interior of compact objects. The natural examples of anisotropy are abundant; here we quote electromagnetic and fermionic fields in neutron stars \cite{sawyer/1973} and superfluidity \cite{carter/1998}. In fact, since the pioneering Reference 16, there has been an considerable number of works \cite{mak/2003,mak/2002,hernandez/2004,maurya/2019,mardan/2019} devoted to the study of anisotropic spherically symmetric static stellar configurations. Let us finally mention that the origin of local anisotropy was investigated in the review of Ref.22 and it was shown that a possible source may be viscosity (see also Ref. 23). Anisotropic models may be important to model very massive compact stars, but we shall see that the class of self-bound versions run into trouble to accommodate radii.

\section{Self-bound matter and stellar models}

The physics of matter at ultra-high densities can be studied in principle with the aid of \textit{Quantum Chromodynamics Theory} (QCD). For large temperatures and densities, QCD matter is \textit{asymptotically free}, the actual region of the phase diagram in which neutron stars or strange stars (SSs) reside present a high uncertainty in the matter behavior \cite{QCD}.

More than 30 years ago, Witten \cite{c} considered the possibility of a self-bound version of quark matter, made stable by the presence of the $s$ quark, elaborating on the previous works of Bodmer \cite{Bod}, Terazawa \cite{Tera} and Itoh \cite{Itoh}. Because of this feature the proposal is known as {\it strange matter}. In this sense, an exotic state of deconfined quarks, could be the true ground state of hadronic matter, not $^{56}Fe$, having a lower energy per baryon than ordinary nuclei. It is generally assumed that in its present version, strange matter is composed of roughly equal numbers of \textit{up}, \textit{down} and \textit{strange} quarks, and a small number of electrons to attain the charge neutrality \cite{farhi/1984,bethe/1987}.

A variety of approaches for a model description of strange matter and, in general, self-bound matter were attempted since Witten's work. The MIT bag model with a quasi-linear equation of state (EoS) has been widely used, but Nambu-Jona-Lasinio \cite{Buballa,Efrain}, density-dependent quark masses \cite{ZhangLi} and a few other variants were also considered. More recently it has been established that {\it paired} quark states should be relevant in the dense deconfined phase. It has been shown that the presence of pairing gaps (quark matter in the Color-Flavor Locked state, neglecting states at intermediate densities for simplicity) actually {\it enhance} the possible stability of the quark matter phase, as discussed by Lugones and Horvath \cite{e,f} because the system's energy is lowered by the negative contribution to the energy from the introduction of a pairing gap. A general parametric study of this possibility has been discussed by Alford and collaborators \cite{Mark}.

A suitable EoS including pairing effects due to the CFL, discussed in Ref. 37, reads

\begin{equation}
\label{eq:t}
    P_r = \frac{1}{3}\rho + \frac{2\psi}{\pi}\rho^{1/2} - \left( \frac{3\psi^2}{\pi^2} + \frac{4}{3}B \right),
\end{equation}
where $\psi$ is defined by
\begin{equation}
\label{eq:y}
    \psi = -\frac{m_s^2}{6}+\frac{2\Delta^2}{3},
\end{equation}
 with $m_s$ being the mass of the strange quark, $\Delta$ is the pairing gap and $B$ is the bag constant. If we assume $m_s \rightarrow 0$ and non-interacting quarks, a \textit{MIT bag model}-like EoS is restored \cite{AFO}.

Other treatments of the quark matter have been presented, and their relevance to the self-boundedness problem addressed. We shall address the model by Franzon et al. \cite{Bruno} applied to the stellar sequences below.

As we have previously mentioned, in our calculations below we shall consider three different quark matter models with and without anisotropy in the interior pressure. Those are NJL Color-flavor locked EoS with vector interactions, Mean Field Theory of QCD (MFTQCD) and Thirukkanesh-Ragel-Malaver {\it ansatz} exact anisotropic models.

\subsection{NJL Color-flavor locked equation of state with vector interactions model}
\qquad

Motivated by the introduction of strange matter, Ferrer \cite{Efrain} discussed a model based on the Nambu-Jona-Lasinio approach, in which vector interactions and gluon components were introduced. This work shows that, even without a gluonic component, NJL EoS can be used to reach high maximum masses in the corresponding sequences provided the ratio of the (repulsive) vector channel to the quark-antiquark $G_{V}/G_{S}$ is high enough. One advantage of this approach is that the gap equation in the CFL phase is calculated in a self-consistent way. In addition, it is well-known that a quantity analogous to a bag constant can also be obtained \cite{Buballa}. The results of Ref.32 for the calculated $B_{0} = 57.3 MeV/fm^{3}$ (note that this quantity numerically coincides with a "MIT bag constant" $B$, but it is determined self-consistently) and $G_{V}/G_{S} = 0.5$ were employed to plot the curve in Fig. 1 below. The introduction of a gluonic component elevates even more the value of the maximum mass for this fixed set $(B_{0}, G_{V}/G_{S})$.

\subsection{MFTQCD (Mean Field Theory of QCD)}

The microphysical model of cold quark matter presented in Ref.39 starts with a separation of ``soft'' and ``hard'' momentum components, and after an analytical calculation one is left with an improved version of a Bag model, termed as MFTQCD. The EoS depends on two quantities: one is $\xi$, the ratio of the coupling to the dynamical gluon mass generated by interactions, and the other is related to the expectation values of the gluon condensates, identified as a ``vacuum energy constant'' and denoted as $B_{QCD}$. Given reasonable values for both quantities, a region in the $\xi - B_{QCD}$ plane in which the self-boundedness condition, namely that the energy per baryon number unit is $\leq m_{n}$ (or, more precisely, the mass of $^{56} Fe/56$) is satisfied. The pairing of quarks was not considered in Franzon et al.\cite{Bruno}, although it can be easily introduced. A stellar sequence generated using the set $\xi = 0.003658 \, MeV^{-1}$ and $B_{QCD} = 62 MeV/fm^{3}$ unpaired quark matter achieved the highest maximum TOV mass, and these are the values shown in Fig.1.

\subsection{Thirukkanesh-Ragel-Malaver {\it ansatz} exact anisotropic models}

\label{subsec:Malaver}
Finally we plot the result for two values of the parametric vacuum energy $B$ that generated anisotropic stellar models based on the Thirukkanesh-Ragel-Malaver {\it ansatz} as proposed by Thirukkanesh and Ragel \cite{k}  and employed by Malaver \cite{l} for the spacetime metric. With the EoS of Eq.(1), Rocha et al.\cite{Gardel} were able to integrate analytically the full problem and obtain stellar sequences with very
high mass, as shown in Fig. 1 for these two values of the vacuum energy density and fixed strange quark mass
$m_{s}$ and pairing gap $\Delta$. The exact solutions display a feature which is not uncommon in a variety of anisotropic stellar models, and it is worth to mention that a general procedure to generate all static, spherically symmetric solutions has been recently presented \cite{aniso}. We believe that these kind of solutions will attract much attention in the near future.

\begin{figure}[htbp!]
\centering
\includegraphics[width=4in]{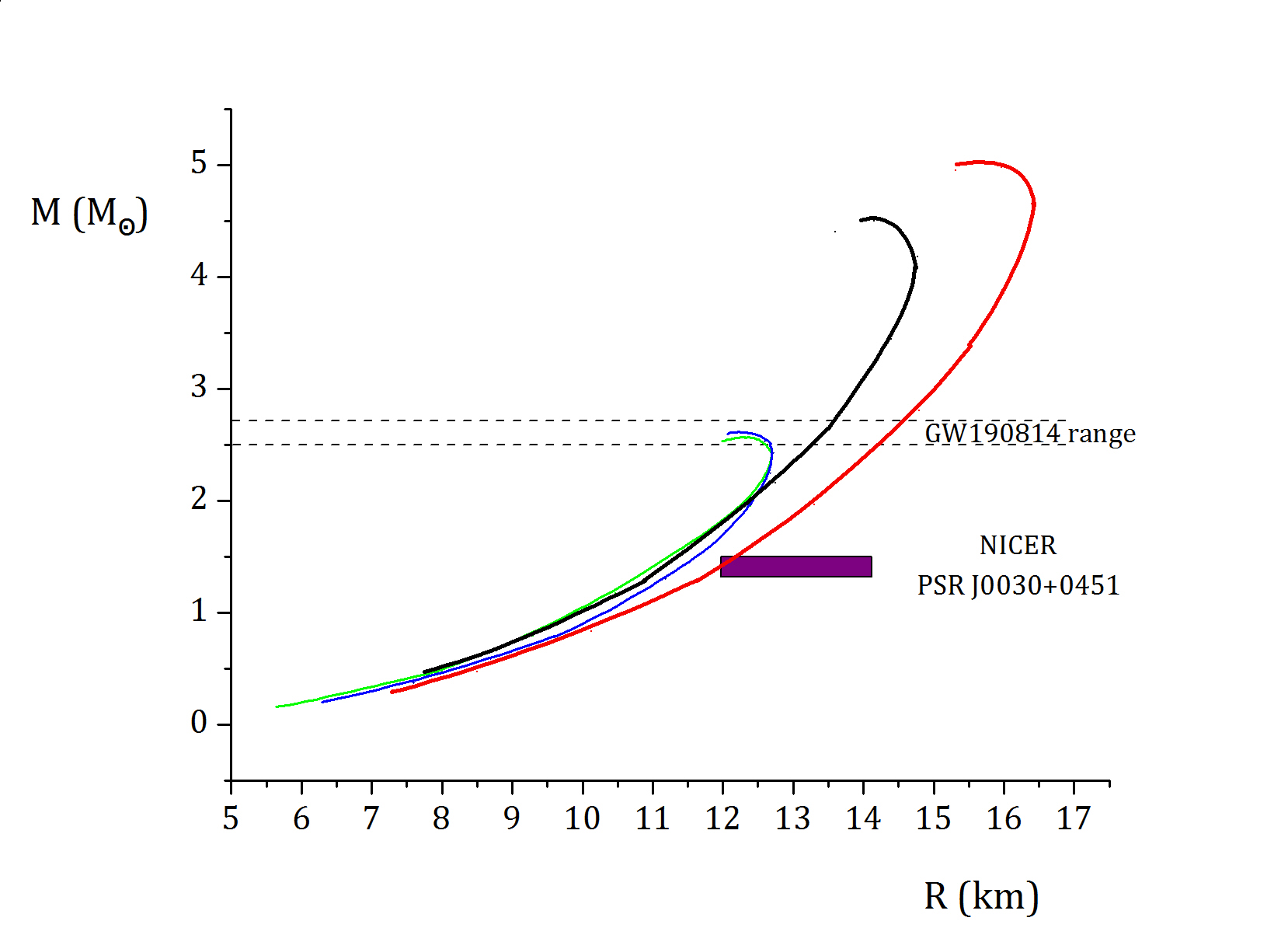}
\caption{\label{fig:k} Mass-radius relation of the theoretical models. The curve generated using the MFTQCD quark EoS, with $\xi = 0.003658 \, MeV^{-1}$ and $B_{QCD} = 62 MeV/fm^{3}$ appears in green. The NJL CFL+vector interactions EoS with $B_{0} = 57.3 MeV/fm^{3}$ and $G_{V}/G_{S} = 0.5$ is the blue curve. The two anisotropic models calculated using the Thirukkanesh-Ragel-Malaver ansatz, both with $\Delta = 100~\mathrm{MeV}$ and $m_s = 150~\mathrm{MeV}$ are the red curve (corresponding to $B = 57.5~\mathrm{MeV/fm^3}$) and the black one ($B = 70~\mathrm{MeV/fm^3}$). The range of masses reported for the lighter object in the merge GW190814 is indicated by the dashed horizontal lines, and the region of the measured values for PSR J0030+0451 with the filled rectangle.}
\end{figure}

\section{Other models for the massive object and implications for the equation of state and stellar physics}

Our considerations of some examples of microphysical models that can be used to model a high-mass
compact star within the class of quark-based models are not exhaustive, other proposals have been published along these lines. There other effects that may allow an effective $M_{max} \geq 2.5 M_{\odot}$. The examples of Most et al. \cite{Most} and Zhang and Li \cite{Nai-Bo}, among others,
show that a rapidly rotating compact star could be involved. As a general feature, the analysis known as I-Love-Q has been often invoked to state that the amount of mass increase due to rotation is $\sim 20 \%$ for almost all equations of state \cite{chin}, although this figure actually depends on the kind of rotating and exact scaling may be broken \cite{Kostas}. Even though there is no clear indication for a rapidly rotation object in the data, this possibility for increasing the mass over the static limit can not be ruled out. Dexheimer et al.\cite{VD} suggest that the rapid rotation does {\it not} preclude an exotic core composition either, while Li, Sedrakian and Weber \cite{Armen} found that $\Delta$-resonance admixed hypernuclear constitution is in difficulty for producing such a massive configuration, even if maximally rotating at the Keplerian rate, and the object is likely to be a black hole. However, they have also shown that if no new degrees of freedom appear to ``hyperonize'' matter, the stiffness would be enough for a compact star interpretation of the object \cite{Sedra}. The black hole interpretation is also favored by Fattoyev et al. \cite{Fatto} after showing that a large stiffening of the equation of state within a covariant density functional theory approach makes difficult to satisfy constraints from heavy ion collisions and deformability of the lower-mass neutron stars obtained from the event GW170817.

Finally, Tsokaros, Ruiz and Shapiro \cite{soco} have argued that the analysis of other systems do not support the rapid rotation idea, and therefore the equation of state is the key element that must be very stiff if the lighter object in the merge GW190814 happens to be a compact star and not a black hole. However, we find very suggestive that a population analysis by the LIGO/Virgo Collaboration found a ``gap'' in the merging population \cite{LIGOpop} analogous to the one suggested in the local environment, namely the absence of masses in the range $\sim 2.5 - 6 M_{\odot}$, which puts the mysterious objects on the ``neutron star'' side. The black hole interpretation, however, would imply a gap among their distribution of masses, suggesting that the ``low-end'' follow a different
formation path (see below), unless the sample is very incomplete indeed.

\section{Discussion}
\qquad
We have presented a few representative examples of self-bound stellar models, two sequences of the isotropic pressure class with different EoS, and two anisotropic sequences of the same model corresponding to different sets of parameters of one model, with the aim of showing a general feature that became likely important for an explanation of a $\geq 2.5 M_{\odot}$ object in the event GW190814. Even though the latter is {\it not} confirmed to be a compact star, and may well be a black hole, there are arguments to believe that we should be prepared to find an explanation for a high mass, as suggested by other measurements in binaries \cite{SciChi}. In fact, this possibility has already been considered by Bombaci et al. \cite{Ignazio} and Wu et al. \cite{Xu} among others.

Versions of paired NJL CFL strange matter with vector interactions, as well as a MFTQCD without pairing were considered as isotropic examples. The first covered wide range of the parameters $B$, $m_s$ and $\Delta$, and the MFTQCD one was selected from the set of self-bound versions, as determined by the parameter $\xi$ to comply with a high maximum mass. While the anisotropic models obtained within the Thirukkanesh-Ragel-Malaver {\it ansatz} are exact within the quasi-linear EoS, the isotropic solutions were calculated numerically. None of these solutions introduced ultra-strong magnetic fields or any other ingredient.

An interesting general feature related to the behavior of the solutions is that up to about $\sim 1 M_{\odot}$, all self-bound stars are essentially Newtonian, and their radius for a given mass is just $R= {\bigl( \frac{3M}{4\pi \rho} \bigr)}^{1/3}$ to a good approximation. General relativistic effects make the radii at higher masses to be {\it smaller} than their Newtonian values (i.e. the $M-R$ curves like the ones in Fig. 1 are above their Newtonian equivalents). Therefore, the agreement with the measured data is worse than it would be
when $M \sim 1.5 M_{\odot}$, due to the General Relativity effects. Of course, this is a relatively minor effect when compared to the essential feature shared by the models: $R \rightarrow 0$ when $M \rightarrow 0$ for all self-bound stellar sequences, in sharp contrast with conventional hadronic models which have the opposite tendency, facilitating the interpretation of a relatively large radius for PSR J0030+0451 \cite{Col}. An exception within quark-based models has been recently presented by Zhang and Mann \cite{ZM}, and Roupas, Panotopulos and Lopes \cite{RPL} achieve both the mass and radius constraints but for extreme values of the parameters, as in Lugones and Horvath \cite{LugH}.

This situation may have different solutions. The first is that any of the observed quantities is accurate
but not precise, in other words, that either the mass of the object in GW190814 is overestimated (although a $\sim 2.5 M_{\odot}$ maximum mass was shown to stem from the observed distribution of binaries \cite{Alsing}), or the radius of the lower-mass PSR J0030+0451 is smaller, and thus the tension would be alleviated. The second solution is that both objects do {\it not} belong to the same class, in the sense that only very high masses contain exotic matter, but not the lighter ones. The third solution is, if both objects do belong to the same class, that self-bound models are not a good explanation of both simultaneously. A fourth solution is very simple: the $2.5 M_{\odot}$ was not a compact star, but rather a low-mass black hole (Lu, Beniamini and Bonnerot \cite{CalTech} favor this interpretation within a triple-system origin). Our discussion in Ref. 5 points out that at least a class of known binaries may be expected to produce such black holes, although the system in the merger GW190814 does not belong in any sense or could have originated in a ``spider'' binary, because the most massive component is surely a massive black hole. Other forms to produce low-mass black holes have been recently discussed (Liu and Lai\cite{LiuLai} 2020, Safarzadeh and Loeb\cite{SLoeb} 2020). Ultimately, the answer to the existence of absence of a ``mass gap'' is an important problem that must be solved gathering and analyzing empirical data.

It is remarkable that anisotropy is known to give a significant contribution to the increase in the maximum mass of compact objects. Recall that Fig. 1 shows that anisotropic models can reach maximum masses as high as $\sim 5 \, M_\odot$.
In fact, one of the earliest works by Heintzmann and Hillebrandt \cite{heintzmann/1974} has shown that anisotropy is a quite promising mechanism to increase the maximum masses of neutron stars, and for an arbitrarily large anisotropy there is no upper limit even when radial stability analysis is taken into account. Anisotropy yields higher maximum masses once the effective pressure inside stars is high enough to make the star to sustain more mass against gravitational collapse. It has also being pointed out in \cite{heintzmann/1974} that this is the case when $P_\perp>P_r$. On the other hand, if $P_r>P_\perp$ the maximum mass is expected to decrease. A very relevant discussion has been also presented by Bowers and Liang \cite{bowers/1974}.

Finally, it is important to state that a confirmation of the compact star character of the puzzling object in GW190814 would automatically establish an absolute upper limit to the central density $\rho_{c}$ of any compact star, as shown by Lattimer and Prakash \cite{Jim}. The Tolman IV and Tolman VII exact solutions provide an envelope for all stellar models, including exotics, which for the former reads

\begin{equation}
    \rho_{c} \simeq 1.56 \times 10^{16} {(M_{\odot}/M)}^{2} g \, cm^{-3}
\end{equation}
which is deemed appropriate for microphysical quark models in a broad sense, and yields immediately
$\rho_{c} \simeq 2.18-2.5 \times 10^{15} g \, cm^{-3}$ for the range of the determined mass.

\section{Acknowledgments}
\qquad
J.E. Horvath acknowledge the Fundac\~ao de Amparo
\`a Pesquisa do Estado de S\~ao Paulo and CNPq Federal Agency for partial financial support.
P.H.R.S. Moraes would like to thank CAPES for financial support. An anonymous referee is acknowledged
for providing additional references and observations that helped us to improve the final version.



\end{document}